\documentclass[aps,prc,reprint,showpacs,groupedaddress,onecolumn]{revtex4}
\usepackage{bm}% bold math
\usepackage{amsmath}
\usepackage{amsfonts}
\usepackage{amscd} 
\usepackage{graphicx}
\usepackage{color}

\newcommand{\weight}{\frac{e^{-\sqrt{\bar{p}^{2}+x}}}{\sqrt{\bar{p}^{2}+x}}\rho_{1}(\sqrt{x})}
\def\beq{\begin{equation}}
\def\eeq{\end{equation}}
\def\xf{\tilde{\mathbf{x}}}
\def\yf{\tilde{\mathbf{y}}}

\def\pf{\tilde{\mathbf{p}}}
\def\kf{\tilde{\mathbf{k}}}
\def\qf{\tilde{\mathbf{q}}}
\def\ff{\tilde{{f}}}
\def\gf{\tilde{{g}}}

\def\x{{\sf x}}

\def\y{{\sf y}}

\def\a{{\sf a}}

\begin{document}

\title{Scattering asymptotic conditions in Euclidean relativistic quantum theory}

\author{Gordon Aiello}
\affiliation{Department of Physics and Astronomy, The University of
Iowa, Iowa City, IA 52242, USA}

\author{W.~N.~Polyzou}
\affiliation{Department of Physics and Astronomy, The University of
Iowa, Iowa City, IA 52242, USA}

\date{\today}


\begin{abstract}

  We discuss the formulation of the scattering asymptotic condition as
  a strong limit in Euclidean quantum theories satisfying the
  Osterwalder-Schrader axioms.  When used with the invariance
  principle this provides a constructive method to compute scattering
  observables directly in the Euclidean formulation of the theory,
  without an explicit analytic continuation.

\end{abstract}

\maketitle

\section{Introduction}

The purpose of this paper is to argue that it is possible to calculate
scattering observables directly in the Euclidean representation of
quantum field theory without analytic continuation.  The essential
observation, which is a consequence of the Osterwalder Schrader
reconstruction theorem \cite{Osterwalder:1973dx}, is that there is a
representation of the physical Hilbert space directly in terms of the
Euclidean Green functions without analytic continuation.  There is
also a representation of the Poincar\'e Lie algebra on this space.
This defines a relativistic quantum theory.  Cluster properties of the
Schwinger functions suggest that it should be possible to formulate
scattering problems directly in this representation.

In (1958) Schwinger \cite{Schwinger:pna} argued that as a result of
the spectral condition that time-ordered Green functions had analytic
continuations to Euclidean space-time variables.  The analytically
continued functions satisfy Schwinger-Dyson equations and are moments
of Euclidean path integrals.

The Osterwalder-Schrader axioms
\cite{Osterwalder:1973dx}\cite{glimm:1981} define conditions on a
collection of Euclidean Green functions (Schwinger functions) that
allow the reconstruction of a relativistic quantum theory.  The
Schwinger functions are formally defined as moments of a Euclidean
path integral \cite{Frohlich:1983kp},
\beq
S_n (\x_1, \cdots, \x_n ) =
{\int D[\phi] e^{- A[\phi]} \phi(\x_1) \cdots \phi (\x_n) \over
\int D[\phi'] e^{- A[\phi']}}.  
\label{a.1}
\eeq
The collection of Schwinger functions are Euclidean invariant and
satisfy a condition, called reflection positivity.  Reflection
positivity is used to construct a representation of the Hilbert space
of the field theory.  On this representation of the Hilbert space there
are explicit expressions for the Poincar\'e generators as differential
operators.  The remarkable property is that in this representation the
Hilbert space inner product is expressed directly in terms of the
Euclidean variables, without an explicit analytic continuation.

Given a representation of the Hilbert space and a set of Poincar\'e
generators that satisfy cluster properties it should be possible to
calculate quantum mechanical scattering observables.  Cluster
properties of the Poincar\'e generators in this representation is a
consequence of cluster properties of the Schwinger functions.

In this work we discuss the direct construction of scattering
observables in this Euclidean framework.  No analytic continuation is
used.  The input is a set of Schwinger functions.  These contain the
dynamics and need to be computed as input.  Their computation is
easier than computing the corresponding Minkowski Green functions.  In
this work we show that scattering observables can be expressed as a
quadratic form with a kernel constructed out of Schwinger functions.
This avoids the need for analytic continuation or the need to solve
singular integral equations.

We formulate the scattering problem using the two-Hilbert space
formulation of time-dependent scattering\cite{simon}\cite{baumgartl:1983}.  
This separates the internal
structure of the asymptotic states from their space-time properties.

Below we list the steps that are needed to construct 
transition matrix elements.  These will be discussed in more
detail in the body of this paper.  While some of these steps
have been discussed in previous work, we include all of them
to make this work self contained.
\begin{itemize}

\item [1] An asymptotic Hilbert space is defined as the direct
  sum of tensor products of irreducible representations of the
  Poincar\'e group.  The mass and spin of the particles in the target,
  beam, and measured in the detectors determine the mass and spin
  labels of the irreducible representations.

\item [2] Mappings from the asymptotic Hilbert space to the Euclidean
  representation of the physical Hilbert space are constructed.  These
  mappings add the internal structure of the asymptotic particles to
  the space-time properties, including particle cloud effects. They
  also control the joint energy-momentum spectrum of the asymptotic
  states, which leads to strong limits.

\item [3] The scattering asymptotic condition is formulated and 
  sufficient conditions on the connected parts of the Schwinger
  functions for the existence of the channel scattering operators
  are given.

\item [4] The invariance principle is used to replace $H$ and $H_0$ by
  $-e^{-\beta H}$ and $-e^{-\beta H_0}$ where $\beta$ is a parameter
  that defines the energy scale.  This allows for a direct Euclidean
  evaluation of $S$ matrix elements.
  
\item [5] The computation of $S$-matrix elements with
  narrow wave packets in momentum is discussed.  Sharp momentum
  transition matrix elements can be factored out of these $S$ matrix elements
  provided the transition matrix elements vary slowly on the support
  of these wave packets.

\end{itemize}

Several aspects of this program were discussed in previous work
\cite{Wessels:2003af}\cite{Kopp:2011vv}\cite{Polyzou:2013nga}. 
In \cite{Kopp:2011vv} a solvable quantum mechanical model
was used to demonstrate that sharp-momentum transition matrix elements
could be accurately computed for a range of energies (from 50
MeV to 2 GeV) by using the invariance principle to replace $H$ by
$-e^{-\beta H}$ in the expression for the M{\o}ller wave operators.
In these calculations $\beta$ is a parameter that sets the working
energy scale.  This computation utilized a uniformly convergent
polynomial approximation of $e^{inx}$ for $x \in [0,1]$.  This
approximation is not necessary in the Euclidean framework.

In \cite{Polyzou:2013nga} the implementation of this method in the
Euclidean framework was discussed.  A class of model
reflection-positive four-point Schwinger functions was given.  The
existence of wave operators for these model Schwinger functions was
established using a generalization of Cook's method \cite{Cook:1957}.
This assumed that the operators that map the
asymptotic Hilbert space to the physical Hilbert space could be
constructed.

The primary purpose of this paper is to complete this program by
showing how to construct the mappings from the asymptotic Hilbert
space to the physical Hilbert space that are needed to establish the 
existence of wave operators using Cook's method.
We also show how the Euclidean representation of the
Hilbert space can be used to eliminate the need for the polynomial
approximation to $e^{-\beta H}$ that was utilized in \cite{Kopp:2011vv}.

The final result is an expression for scattering observables as a 
quadrature using the Schwinger functions and narrow wave packets
as input.

The mapping from the asymptotic Hilbert space to the Euclidean
representation of the Hilbert space is the Euclidean analog of a
Haag-Ruelle quasi-local field operator
\cite{Haag:1958vt}\cite{Brenig:1959}\cite{Ruelle:1962}\cite{simon}\cite{jost}.
This mapping controls the four-momentum spectrum of the asymptotic
states, which is needed to isolate asymptotic states with different
mass and the same energy.  This can be done in the Euclidean
representation because we have explicit representations of the four
momentum operators.  The result is that the scattering asymptotic
condition can be formulated as a strong limit, like it is in
non-relativistic quantum mechanics.

In section two the Euclidean representation of the physical Hilbert
space given by Osterwalder and Schrader is defined.  In section three
expressions for the Poincar\'e generators in this representation of
the Hilbert space are given.  The discussion in this paper is limited
to spinless particles.  Spin does not introduce any additional
complications that impact the formulation of the scattering problem.
The treatment of particles with spin is discussed in
\cite{Polyzou:2013nga}.  In section four the results of
these sections are illustrated using the Lehmann representation of a
two-point Schwinger function.  Methods for isolating the discrete part
of the Lehmann weight of these functions are central to the
formulation of the scattering asymptotic condition. In section five
the abstract two-Hilbert space formulation of the scattering
asymptotic condition is discussed.  This requires ``Euclidean
Haag-Ruelle'' injection operators that map the asymptotic Hilbert
space to the physical Hilbert space.  The construction of these operators
is given in section six.  In section seven the construction of
sharp momentum transition matrix elements is discussed.  It is shown
how properties of the Euclidean representation of the Hilbert space
can be used to avoid the polynomial approximation used in
\cite{Kopp:2011vv}.  Explicit formulas that express $S$-matrix
elements directly in terms of the Euclidean Schwinger functions
without analytic continuation are given.  The appendix contains the
main technical results of this paper.  It shows that orthogonal
polynomials in the mass squared operator are complete in this
representation of the Hilbert space.  This is needed to ensure that
the injection operators that are used to formulate the asymptotic
condition have positive relative-time support,  so their
range is in the Euclidean representation of the Hilbert
space.  The material in sections one-five is not new;  it is included
to make the article as self-contained as possible and limited to 
material that is essential to the new material.  The new material
is in sections six, seven and the appendix.

\section{Hilbert space} 

In the Euclidean framework Hilbert space vectors are represented by 
sequences of Schwartz test functions of the form
\beq
f (x)  :=  ( f_1 (\x_{11}), f_2 (\x_{21},\x_{22}),\cdots )
\label{b.1}
\eeq
where the individual functions have support restrictions
\beq
f_n (\x_{n1}, \x_{n2},\cdots , \x_{nn}) =0
\qquad
\mbox{unless} 
\qquad
0 < \x_{n1}^0 <  \x_{n2}^0  <\cdots <  \x_{nn} .
\label{b.2}
\eeq
Sequences of functions $\{f_n \}$ satisfying this support condition
are called positive relative time functions.  This is a linear
subspace of the space of sequences of Schwartz test functions in Euclidean
variables.

The Euclidean time reflection operator $\Theta$ is defined on these
sequences. It changes the sign of the Euclidean times in each $f_n$: 
\beq
\Theta f (x) = f(\theta x) =
\left (f_1 (\theta \x_{11}), f_2 (\theta \x_{21}, \theta \x_{22}), \cdots
\right )
\label{b.3}
\eeq
where
\beq
\theta \x = \theta (x_{ij}^0,\mathbf{x}_{ij}) =
(-x_{ij}^0,\mathbf{x}_{ij}).
\label{b.4}
\eeq
The Hilbert space inner product of $f$ with $g$  is defined by 
\[
\langle f \vert g \rangle =
\]
%\[
%\sum_{mn} \int d^{4n}x d^{4m}y
%f_n^* (\theta \x_{n1}, \theta\x_{n2},\cdots ,\theta \x_{nn})
%S_{m+n}
%(\x_{nn}, \cdots , \x_{1n},\y_{1m} , \cdots , \y_{mm})
%g_m (\y_{m1}, \y_{m2},\cdots , \y_{mm}) =
%\]
\beq
\sum_{mn} \int d^{4n}x d^{4m}y
f_n^* ( \x_{n1}, \x_{n2},\cdots , \x_{nn})
S_{m+n}
(\theta \x_{nn}, \cdots , \theta \x_{1n},\y_{1m} , \cdots , \y_{mm})
g_m (\y_{m1}, \y_{m2},\cdots , \y_{mm}) .
\label{b.5}
\eeq
It is instructive to write this using the following short-hand notation
\beq
\langle f \vert g \rangle = 
%(\Theta \Pi f ,S \Pi g) =
(f , \Pi \Theta  S \Pi g) 
\label{b.6}
\eeq
where $\Pi$ is the projection of the subspace of Euclidean
test functions with positive relative time support and
$\Theta$ is the Euclidean time reversal operator.

Reflection positivity is the condition that this space does not have
negative norm states:
\beq
\langle f \vert f \rangle = (f ,\Pi \Theta S \Pi f) \geq 0 .
\label{b.7}
\eeq
Reflection positivity makes (\ref{b.5}) into a Hilbert space inner
product.  There are sequences with zero
norm.  The physical Hilbert space is obtained by identifying sequences
whose difference has zero norm.  This space is made complete by
identifying Cauchy sequences with vectors in this space.  While the
Euclidean time supports must be disjoint, the order in (\ref{b.2})
does not matter because the Schwinger functions of a local field
theory are symmetric \cite{Osterwalder:1973dx}, so the arguments can
be relabeled so (\ref{b.2}) is satisfied.

Reflection positivity is not automatic; it is a property of acceptable
Schwinger functions.  By Euclidean invariance, reflection positivity
must hold for any choice of the Euclidean time axis.  In addition,
without the support restriction, functions that are odd with respect
to Euclidean time reflection will have negative norm, so the support
condition, or an alternative restriction is necessary for positivity.

In what follows we assume that the collection of Schwinger functions
are reflection positive.  This condition needs to be verified in
models or approximations.

\section{The Poincar\'e Lie Algebra}

The Euclidean time reflection breaks the Euclidean
invariance of (\ref{b.5}).  As a result the group of real Euclidean 
transformations on sequences of Euclidean test functions
\beq
f_n (\x_{n1}, \x_{n2},\cdots , \x_{nn}) \to
f_n^{\prime } (\x_{n1}, \x_{n2},\cdots , \x_{nn}) =
f_n (O \x_{n1}-a , O\x_{n2}-a ,\cdots , O\x_{nn}-a ), 
\label{c.1}
\eeq 
where $O \in O(4)$ and $a$ is a Euclidean four vector, 
becomes a subgroup of the complex Poincar\'e group with respect to the
inner product (\ref{b.5}).  This is because the covering groups of the 
Lorentz group and $O(4)$ have the same analytic continuation.

The only technical issue is that these complex transformations 
do not generally preserve the positive relative time condition, however, 
Euclidean time translations map positive relative time functions into
positive relative time functions when $a^0\geq 0$.  Similarly
rotations in Euclidean space-time planes are defined for small
angles on a subspace of positive relative-time functions with support
on wedge shaped regions $x^0_{kn} > b \vert \mathbf{x}_{kn}\vert$
where $b$ is a positive constant.  These transformations become
contractive Hermetian semigroups and local symmetric semigroups 
\cite{Klein:1981}
\cite{Klein:1983}
\cite{Frohlich:1983kp} on the
Hilbert space (\ref{b.5}) 
provided the Schwinger functions are
sufficiently regular.  We assume that the collection of Schwinger
functions satisfy these conditions.  These transformations are
associated with imaginary-time time translations and Lorentz boosts
with imaginary rapidity.  The important point is that in both cases
the generators of contractive Hermetian semigroups and local symmetric
semigroups are self-adjoint.

The infinitesimal generators $H$ and $\mathbf{K}$ of time translations
and rotationless boosts, along with the generators $\mathbf{J}$ and
$\mathbf{P}$ of rotations and translations $\mathbf{P}$ have the
following representations as differential operators
\[
H f_n (\x_{n1}, \x_{n2},\cdots , \x_{nn})
=
\sum_{k=1}^n {\partial \over \partial \x_{nk}^0} f_n (\x_{n1}, \x_{n2},\cdots , \x_{nn})
\]
\[
\mathbf{P}  f_n (\x_{n1},\x_{n2},\cdots , \x_{nn})
=
-i \sum_{k=1}^n {\partial \over \partial \mathbf{x}_{nk}} f_n (\x_{n1}, \x_{n2},\cdots ,\x_{nn})
\]
\[
\mathbf{J} f_n ( \x_{n1}, \x_{n2},\cdots ,
\x_{nn}) = -i \sum_{k=1}^n \mathbf{x}_{nk} \times {\partial \over
\partial \mathbf{x}_{nk}} f_n (\x_{n1},
\x_{n2},\cdots ,\x_{nn})
\]
\beq
\mathbf{K} f_n ( \x_{n1}, \x_{n2},\cdots ,
\x_{nn}) = \sum_{k=1}^n (\mathbf{x}_{nk} \times {\partial \over
\partial {x}^0_{nk}}-  {x}^0_{nk}  {\partial \over
\partial \mathbf{x}_{nk}}) f_n ( \x_{n1},
\x_{n2},\cdots ,\theta \x_{nn}).
\label{c.2}
\eeq
Direct computation shows that these operators are Hermetian and
satisfy the commutation relations of the Poincar\'e Lie algebra. 
It can be shown that the Hamiltonian is bounded from below \cite{glimm:1981}.
Both the boost generators and Hamiltonian do not have the usual factor of
$i$.  This is because they are linear in $\x^0_n$ or ${\partial \over
\partial x^0_m}$ so the usual sign change from complex conjugation is
replaced by the Euclidean time reflection.

In the Euclidean framework the Poincar\'e generators are simple
differential operators; the dynamical information is contained in the
Schwinger functions.  A collection of reflection positive Schwinger
functions and the expressions for the Poincar\'e generators define the
relativistic quantum theory.

\section{Two-point functions}

The two-point Schwinger function illustrates the relation of the
Euclidean representation of a relativistic quantum theory to the
conventional treatment of a relativistic particle.  The structure of
the two-point Schwinger function is also used to formulate the
scattering asymptotic condition in the Euclidean framework.

The structure of the two-point Schwinger function is given by
its Lehmann representation, which has the form
\beq
S_2(\x-\y) = {1 \over (2 \pi)^4} \int
{d^4 p \rho (m) dm  \over p^2 + m^2 }  e^{i p \cdot (x-y) }
\label{d.1}
\eeq
where $\rho(m)$ is the Lehmann weight.  The support of $\rho(m)$ is on
the mass spectrum of the states that have non-zero matrix elements
with $\phi (x) \vert 0 \rangle$, where $\phi (x)$ is the Minkowski
Heisenberg field and $\vert 0 \rangle$ is the vacuum.  The spectral
condition requires that the support of $\rho(m)$ is for positive
values of $m$.

Using this general two-point Schwinger function to compute the inner 
product (\ref{b.5}) for vectors represented by a 
single function with positive-time support gives
\[
\langle f\vert g \rangle =  
%\int f^*(\theta \x) S_2(\x-\y) g(\y) d^4\x d^4\y=
\int f^*(\x) S_2(\theta \x-\y) g(\y) d^4\x d^4\y=
\]
\[
{1 \over (2 \pi)^4} \int
f^*(\x)
{ e^{i p^0 (-x^0-\y^0) +i \mathbf{p} \cdot (\mathbf{x}-\mathbf{y})}
\rho (m)
\over (p^0)^2 + \mathbf{p}^2 + m^2 }  g(\y)  dm 
d^4 p d^4\x d^4\y=
\]
\beq
\int \chi^*(\mathbf{p}) 
{ \rho (m)  \over 2 \omega_m(\mathbf{p})} dm  d\mathbf{p}
\psi (\mathbf{p}) 
\label{d.1a}
\eeq
where the momentum-space wave functions are
\beq
\chi(\mathbf{p}) =
\int {d\mathbf{x} \over (2 \pi)^{3/2} }
f (\x^0,\mathbf{x}) e^{- \omega_m (\mathbf{p}) \x^0
  - i \mathbf{p} \cdot \mathbf{x} } \qquad 
\phi (\mathbf{p})=
\int {d\mathbf{y} \over (2 \pi)^{3/2} }g (\x^0,\mathbf{x}) e^{- \omega_m (\mathbf{p}) \x^0
- i \mathbf{p} \cdot \mathbf{y} }
\label{d.2}
\eeq
and $\omega_m (\mathbf{p}) =\sqrt{m^2 + \mathbf{p}^2}$ is the energy
of a particle with mass $m$ and momentum $\mathbf{p}$.  Here the
support condition on the Euclidean times along with the time
reflection on the final state allows the $p_0$ integral to be
evaluated using the residue theorem.  There are several important
observations about this expression.

\begin{itemize}

\item [1] If $f(\x) = g(\x)$ then $\langle f \vert f \rangle \geq 0$, so
the Lehmann representation of two-point Schwinger functions is
consistent with reflection positivity.  

\item [2] Eq. (\ref{d.1}) has the form of an ordinary Lorentz invariant inner
product with Lorentz invariant measure $d \mathbf{p}/2 \omega_m
(\mathbf{p})$. This shows how the physical Hilbert
space inner product emerges from this Euclidean quadrature
without analytic continuation.

\item [3] The origin of the mass dependence, which provides the
dynamical relation between energy and momentum, is in the two-point
Schwinger function.

\item [4] The distribution $f(\x) = \delta (\x^0 -c) \tilde{f}
(\mathbf{x})$, $(c>0)$ is square integrable in the 
inner product (\ref{d.1a}) when $\tilde{f}
(\mathbf{x})$ is a Schwartz test function in three variables.  This
is due to the non-trivial kernel in the scalar product. This means
that the Hilbert space of normalizable vectors includes expressions
with delta functions in Euclidean time.  This observation has
useful computational consequences which will be used in section 8.

\item [5] The mass square Casimir operator 
is the four-dimensional Euclidean Laplacian: 
\beq
M^2 = \nabla^2_e : ={\partial^2 \over \partial \x^{02}} + 
\pmb{\nabla}_{\mathbf{x}}^2 
\label{d.4}
\eeq

\item [6] $\rho (m) = \delta (m-m_0)$ is the Lehmann
weight for a free particle of mass $m_0$.

\end{itemize}

The Lehmann weight $\rho(m)$ has the general structure
\beq
\rho(m) = \sum_{k=1}^N z_k \delta (m-m_k)
+ \rho_{ac} (m)
\label{d.5}
\eeq
where
\beq
0 < m_1 < m_2 < \cdots < m_n < \mbox{support} (\rho_{ac}). 
\label{d.6}
\eeq
The support of the continuous part of the Lehmann weight,
$\rho_{ac}$, which is associated with multiparticle states, is not
bounded from above.  The continuous part of the Lehmann weight is 
polynomially bounded \cite{glimm:1981}.  We assume this condition and that
the support of the Lehmann weight also includes discrete masses.

\section{Scattering asymptotic condition}

The scattering asymptotic condition is formulated using a two Hilbert
space framework \cite{Coester:1965zz}\cite{Chandler:1980}\cite{Coester:1982vt}\cite{simon}\cite{baumgartl:1983}.
%The basic formulation is abstract; the starting 
%point is a representation of the Hilbert space and a representation 
%of the Poincar\'e group on that space.
%This can be found in some texts, but it is
%not widely accessible.
Detectors respond to a particle's mass, spin, linear momentum, and
spin polarization.  The internal structure of the particle is of no
consequence to the detector.  The two Hilbert space formulation of
scattering separates the degrees of freedom that are measured
asymptotically from the internal degrees of freedom.

An $N$-particle scattering state asymptotically looks like a direct
product of wave packets in each particle's momentum and magnetic
quantum number.  $N$-particle wave packets span a channel subspace
defined by
\beq
{\cal H}_{\alpha} = \otimes_{i=1}^{n_\alpha} {\cal H}_{m_ij_i}
\label{e.1}
\eeq
where ${\cal H}_{m_ij_i}$ is a mass $m>0$ spin $j$ irreducible
representation space of the Poincar\'e group, associated with
mass and spin of each asymptotic particle in the scattering
channel $\alpha$.  The functions in ${\cal H}_{mj}$ are square 
integrable functions of the three-momentum and magnetic
quantum numbers of a particle of mass $m$ and spin $j$.

A channel $\alpha$ injection operator, $J_\alpha$,
\beq
J_{\alpha} : {\cal H}_{\alpha} \to {\cal H}
\label{e.2}
\eeq
is a mapping from the asymptotic channel $\alpha$ Hilbert space to the
full Hilbert space.  This operator combines the internal structure
degrees of freedom for each asymptotically separated particle in the
channel $\alpha$ with their momentum and spin degrees of freedom.  The
purpose of this paper is to discuss the construction of channel
injection operators in the Euclidean representation.

The asymptotic Hilbert space is defined as the direct sum
of the channel subspaces
\beq
{\cal H}_{\cal A} = \oplus_{\alpha \in {\cal A}}{\cal H}_{\alpha}
\label{e.3}
\eeq
and the two Hilbert space injection operator as the
sum of all of the channel injection operators
\beq
{J}_{\cal A} = \oplus_{\alpha \in {\cal A}}J_{\alpha} .
\label{e.4}
\eeq

There is a natural unitary representation $U_{\cal A}(\Lambda ,a)$ of the
Poincar\'e group on the asymptotic Hilbert space.  It is the 
tensor product of unitary irreducible representations:
\beq
U_{\cal A}(\Lambda ,a) = \oplus_{\alpha \in {\cal A}}
(\otimes_{i \in \alpha} D^{m_ij_i} (\Lambda ,a) ) 
\label{e.5}
\eeq
where the $D^{m_ij_i} (\Lambda ,a)$ are unitary irreducible
representations of the Poincar\'e group for a particle of mass $m_i$
and spin $j_i$:
\beq
D^{m_ij_i} (\Lambda ,a) = \langle (m_i,j_i) \mathbf{p}_i,\mu_i \vert
U_i(\Lambda ,a)\vert (m_i,j_i) \mathbf{p}_i',\mu_i' \rangle .
\label{e.6}
\eeq 
These representations are known analytically \cite{Keister:1991sb}.
In the asymptotic Hilbert space there is no distinction 
between elementary and composite particles.  These distinctions
appear in the injection operator.

A scattering state is a solution of the Schr\"odinger equation
that evolves into a state that asymptotically
looks like a collection of asymptotically separated free particles.  In
the two-Hilbert space formalism the asymptotic condition has the form
\beq
\lim_{t \to \pm \infty}
\Vert U(t,I) \vert \psi_{\pm} \rangle - J_{\cal A} U_{\cal A}(t,I) 
\vert \psi_0 \rangle \Vert =0 .
\label{e.7}
\eeq
Using the unitarity of $U(t,I)$ this can be expressed as
\beq
\lim_{t \to \pm \infty}
\Vert \vert \psi_{\pm} \rangle - U(-t,I) J_{\cal A} U_{\cal A}(t,I) 
\vert \psi_0 \rangle \Vert =0 .
\label{e.8}
\eeq
The existence of this limit depends on properties of time evolution
subgroup of the Poincar\'e group, $U(t,I)$, and the choice of
injection operator, $J_{\cal A}$.  A sufficient condition for the
convergence of this limit is the Cook condition \cite{Cook:1957}, which
in the two-Hilbert space framework is
\beq
\int_a^{\pm \infty} 
\Vert (H J_{\cal A}  - J_{\cal A} H_{\cal A} )
U_{\cal A}(t,I) \vert \psi_0 \rangle \Vert dt < \infty  
\label{e.9}
\eeq
where $a$ is finite.
For this to hold $(H J_{\cal A} - J_{\cal A} H_{\cal A} )$
needs to be short-ranged operator.   This asymptotic condition can be verified
one channel at a time:
\beq
\int_a^{\pm \infty} 
\Vert (H J_{\alpha}  - J_{\alpha} H_{\alpha} )
U_{\alpha}(t,I) \vert \psi_0 \rangle \Vert dt < \infty .
\label{e.10}
\eeq
The discussion above applies to any representation of a quantum
theory.  In the Euclidean representation the integrand in the 
Cook condition can be
expressed in terms of the inner product (\ref{b.5}):
\[
\Vert (H J_{\alpha} - J_{\alpha} H_{\alpha} )
U_{\alpha}(t,I) \vert \psi_0 \rangle \Vert =
\]
\beq
( \psi_0 \vert U_{\alpha}(-t,I) 
(J^{\dagger}_{\alpha}H - H_{\alpha} J^{\dagger}_{\alpha}) 
\Theta  S 
(H J_{\alpha} - J_{\alpha} H_{\alpha} ) U_{\alpha}(t,I)
\vert \psi_0 )^{1/2} .
\label{e.11}
\eeq

\section{Injection operators and one-body solutions}

In this section we discuss the construction of injection operators
that can be used to formulate the scattering asymptotic condition
(\ref{e.7}).
The basic strategy that we employ is a Euclidean reformulation of
Haag-Ruelle \cite{Haag:1958vt}\cite{Brenig:1959}
\cite{Ruelle:1962}\cite{simon} scattering.

One difference with quantum theories of a fixed finite number of
particles and quantum field theory is that while the field theory
Hamiltonian has one-body eigenstates, there are no states
corresponding to $N$ {\it free} particles.  The absence of a subspace
corresponding to $N$ free particles impacts the formulation of the
scattering asymptotic condition.  The important observation is that
the asymptotic condition defines the initial condition of
Schr\"odinger equation at a time when the particles are asymptotically
separated.  The condition requires that at this time the solution
looks like another state of $N$ asymptotically separated particles.
These other states provide labels for different $N$-particle
scattering solutions of the Schr\"odinger equation.  What these other
states look like when the particles are not asymptotically separated is
irrelevant to the formulation of this initial condition, however the
choice of these states impact the labels that distinguish different
scattering solutions.

To formulate a suitable asymptotic condition in the field theory
case Haag and Ruelle construct localized field operators that
create single-particle states out of the vacuum.  For scalar particles
these quasi-local field operators have the form 
\beq
\phi_h (x) = {1 \over (2 \pi)^2} \int h(-p^2) e^{i p \cdot (x -y)} \phi (y)  
d^4p d^4 y
\label{f.1}
\eeq
where $h(-p^2)$ is a smooth function that is $1$ when $p^2=-m^2$
is the mass of the asymptotic particle, and vanishes on the 
rest of the support of the Lehmann weight of the two-point function.   Then they isolate the part of $\phi_h(x)$
that asymptotically behaves like a creation operator for a particle of 
mass $m$:
\beq
a^{\dagger} (f) = i \int \phi_h (x) \stackrel{\leftrightarrow}{\partial_t} 
f(x) d^3 x
\label{f.2}
\eeq
where $f(x)$ is a positive-energy solution of the Klein-Gordan
equation for a particle of mass $m$.  When applied to the vacuum this
operator creates a time-independent single-particle state out of the vacuum.
Products of these operators with different positive-energy Klein Gordon
solutions, $f_i (x_i)$
are time dependent, and represent \cite{simon}
\beq 
U(-t,I) J_{\cal A} U_{\cal A}(t,I) 
\vert \psi_0 \rangle 
\label{f.3}
\eeq
in (\ref{e.8}).  Ruelle demonstrated the existence of the strong
limits ($t \to \pm \infty$) using locality and the assumed existence
of a mass gap.  The existence of the strong limits in the Haag-Ruelle
case is a result of using the quasi-local fields $\phi_h(x)$, rather
than local interpolating fields $\phi (x)$ used in LSZ theory, 
to formulate the 
scattering asymptotic condition.  Ruelle was able to show
that the integrand in (\ref{e.9}) asymptotically behaves like
$t^{-3(n-1)/2}$, \cite{jost} which is sufficient to satisfy the Cook
condition.

In order illustrate the corresponding construction in the Euclidean
representation we focus on the Cook condition for elastic two-particle
scattering.  This can be formulated in terms of the four-point Schwinger
function, $S_4(\x_1, \x_2, \y_2, \y_1)$.

In order to support a scattering theory the 
four-point Schwinger function should have a cluster expansion of the form:
\beq
S_4(\x_1, \x_2, \y_2, \y_1) =
S_2(\x_1,y_1) S_2(\x_2, \y_2 ) + S_c (\x_1, \x_2, \y_2, \y_1)  
\label{f.4}
\eeq
where $S_c (\x_1, \x_2, \y_2, \y_1)$ is translationally invariant, but
otherwise connected.  The two-point functions $S_2(\x_i, \y_i )$ have
a standard Lehmann representations of the form (\ref{d.1}).  In the
physically interesting case the Lehmann weight in $S_2$ has both discrete and
continuous parts.

Equation (\ref{e.7}) has two distinct contributions; 
One from the connected part, $S_c(\x_1, \x_2, \y_2, \y_1)$,
of the four-point function and the other from the disconnected part,
$S_2(\x_1,y_1) S_2(\x_2, \y_2 )$.  

The quantities that
appear in (\ref{e.7}) are:
\beq
H_\alpha = \omega_{m_1}(\mathbf{p}_1) + \omega_{m_2}(\mathbf{p}_2)
\qquad
\omega_{m_1}(\mathbf{p}_1) := \sqrt{m_i^2 + \mathbf{p}_i^2} 
\label{f.5}
\eeq
\beq
U_\alpha (t,I) = 
e^{-i (\omega_{m_1}(\mathbf{p}_1) + \omega_{m_2}(\mathbf{p}_2))t}
\label{f.6}
\eeq
\beq
H = {\partial \over \partial x^{0}_1} + {\partial \over \partial x^{0}_2} .
\label{f.7}
\eeq
The matrix elements of the injection operator $J_\alpha$ have the general structure
$
\langle \x_1 , \x_2 \vert J_\alpha \vert \mathbf{p}_1 , \mathbf{p}_2 \rangle .
$
Using these expressions in 
(\ref{e.11}) gives
\[
\Vert (H J_{\alpha} - J_{\alpha} H_{\alpha} )
U_{\alpha}(t,I) \vert \psi_0 \rangle \Vert^2 =
\]
\[
\int \psi_1^* (\mathbf{p}_1) \psi_2^* (\mathbf{p}_2) 
e^{i  (\omega_{m_1}(\mathbf{p}_1) + \omega_{m_2}(\mathbf{p}_2))t}
(
{\partial \over \partial x^{0}_1} + {\partial \over \partial x^{0}_2}
- 
\omega_{m_1}(\mathbf{p}_1) - \omega_{m_2}(\mathbf{p}_2))
\langle \mathbf{p}_1 , \mathbf{p}_2  \vert J^{\dagger} 
\vert \x_1 , \x_2 \rangle
\times 
\]
\[ 
\left (
S_2(\theta \x_1,y_1) S_2(\theta\x_2, \y_2 ) + S_c (\theta \x_1, \theta \x_2, \y_2, \y_1)
\right ) 
\]
\[
(
{\partial \over \partial y^{0}_1} + {\partial \over \partial y^{0}_2}
- 
\omega_{m_1}(\mathbf{p}_1') - \omega_{m_2}(\mathbf{p}_2'))
\langle \y_1 , \y_2 \vert J  \vert \mathbf{p}_1' , \mathbf{p}_2'  \rangle
\]
\beq
e^{-i  (\omega_{m_1}(\mathbf{p}_1') + \omega_{m_2}(\mathbf{p}_2'))t}
\psi_1 (\mathbf{p}_1') \psi_2 (\mathbf{p}_2') 
d\mathbf{p}_1
d\mathbf{p}_2
d\mathbf{p}_1'
d\mathbf{p}_2'
d^4x_1 d^4 x_2 d^4 y_1 d^4 y_2 .
\label{f.8}
\eeq
We show that for a suitable choice of injection operator $J$ the
contribution to (\ref{f.8}) from the product of two-point functions
vanish.  When this is true the entire contribution to (\ref{f.8}) comes
from the connected part of the four-point function.

To show this we take advantage of the observation that  
a delta function in Euclidean time multiplied by a Schwartz test 
function in space variables is a normalizable vector in the 
Euclidean representation of the Hilbert space.  
 
We consider an injection operator $J$ of the form
\beq
\langle \x_1, \x_2  \vert  J \vert \mathbf{p}_1, \mathbf{p}_2 \rangle
= h_1 (\nabla_1^2) h_2 (\nabla_2^2)
\delta (\x_1^0 - \tau_1 )\delta (\x_2^0 - \tau_2 )
{1 \over (2 \pi)^3} e^{i \mathbf{p}_1\cdot  \mathbf{x}_1
+ i \mathbf{p}_2\cdot  \mathbf{x}_2}
\label{f.9}
\eeq
where $h_i (m^2)$ is a function that is $1$ when
$m^2$ is the square of the asymptotic particles mass and
$0$ on the rest the support of the Lehmann weight.

A few remarks are in order.  First, the Euclidean time support
condition can be satisfied by choosing $0 < \tau_{1} < \tau_{2}$.  The
non-trivial constraint on the $h_i (\nabla_i^2)$ is that these
operators map functions with positive Euclidean time support to
functions with positive Euclidean time support.  The construction of
functions with this property will be discussed in the next section.
For the purpose of this section we assume that the chosen
$h_i(\nabla_i^2)$ preserves the support condition.

With this choice the disconnected part of (\ref{f.8}) becomes
\[
\int \psi_1^* (\mathbf{p}_1) \psi_2^* (\mathbf{p}_2) 
e^{i  (\omega_{m_1}(\mathbf{p}_1) + \omega_{m_2}(\mathbf{p}_2))t}
(
{\partial \over \partial \x^{0}_1} + {\partial \over \partial \x^{0}_2}
- 
\omega_{m_1}(\mathbf{p}_1) - \omega_{m_2}(\mathbf{p}_2))
\times
\]
\[
\langle \mathbf{p}_1 , \mathbf{p}_2  \vert J^{\dagger} 
\vert \x_1 , \x_2 \rangle 
S_2(\theta \x_1,\y_1) S_2(\theta\x_2, \y_2 ) \times
\]
\[
(
{\partial \over \partial y^{0}_1} + {\partial \over \partial y^{0}_2}
- 
\omega_{m_1}(\mathbf{p}_1') - \omega_{m_2}(\mathbf{p}_2'))
\langle \y_1 , \y_2 \vert J  \vert \mathbf{p}_1' , \mathbf{p}_2'  
\rangle 
\times
\]
\[
e^{-i  (\omega_{m_1}(\mathbf{p}_1') + \omega_{m_2}(\mathbf{p}_2'))t}
\psi_1 (\mathbf{p}_1') \psi_2 (\mathbf{p}_2') 
d\mathbf{p}_1
d\mathbf{p}_2
d\mathbf{p}_1'
d\mathbf{p}_2'
d^4x_1 d^4 x_2 d^4 y_1 d^4 y_2  =
\]
\[
\int \psi_1^* (\mathbf{p}_1) \psi_2^* (\mathbf{p}_2) 
e^{i  (\omega_{m_1}(\mathbf{p}_1) + \omega_{m_2}(\mathbf{p}_2))t}
(
{\partial \over \partial \x^{0}_1} + {\partial \over \partial \x^{0}_2}
- 
\omega_{m_1}(\mathbf{p}_1) - \omega_{m_2}(\mathbf{p}_2)) \times
\]
\[
h_1 (\nabla_{\x1}^2) h_2 (\nabla_{\x2}^2)
\delta (\x_1^0 - \tau_1 )\delta (\x_2^0 - \tau_2 )
{1 \over (2 \pi)^3} e^{-i \mathbf{p}_1\cdot  \mathbf{x}_1
-i \mathbf{p}_2\cdot  \mathbf{x}_2}
\times 
\]
\[ 
{1 \over (2 \pi)^6} {d\mathbf{k}_1 d\mathbf{k}_2 \rho_1 (m) \rho_2' (m') dm dm'
\over 4 \omega_{m} (\mathbf{k}_1)\omega_{m'} (\mathbf{k}_2)}
e^{-\omega_{m} (\mathbf{k}_1)(\x_1^0 + \y_1^0)-   
\omega_{m'} (\mathbf{k}_2)(\x_2^0-\y_1^0)}
e^{i \mathbf{k}_1 \cdot (\mathbf{x}_1 - \mathbf{y}_1) +
i \mathbf{k}_2 \cdot (\mathbf{x}_2 - \mathbf{y}_2)}
\]
\[
(
{\partial \over \partial y^{0}_1} + {\partial \over \partial y^{0}_2}
- 
\omega_{m}(\mathbf{p}_1') - \omega_{m'}(\mathbf{p}_2'))
\times
\]
\[
h_1 (\nabla_{\y1}^2) h_2 (\nabla_{\y2}^2)
\delta (\y_1^0 - \tau_1 )\delta (\y_2^0 - \tau_2 )
{1 \over (2 \pi)^3} e^{i \mathbf{p}_1'\cdot  \mathbf{y}_1
+ i \mathbf{p}_2'\cdot  \mathbf{y}_2}
\times
\]
\[
e^{-i  (\omega_{m_1}(\mathbf{p}_1') + \omega_{m_2}(\mathbf{p}_2'))t}
\psi_1 (\mathbf{p}_1') \psi_2 (\mathbf{p}_2') 
d\mathbf{p}_1
d\mathbf{p}_2
d\mathbf{p}_1'
d\mathbf{p}_2'
x^4x_1 d^4 x^2 d^4 y_1 d^4 y_2 =
\]
\[
\int \psi_1^* (\mathbf{p}_1) \psi_2^* (\mathbf{p}_2) 
(
\omega_{m}(\mathbf{p}_1) - \omega_{m'}(\mathbf{p}_2) - 
\omega_{m_1}(\mathbf{p}_1) - \omega_{m_2}(\mathbf{p}_2)) \times
h_1 (m^2) h_2 (m^{\prime 2})
\times 
\]
\[ 
{d\mathbf{p}_1 d\mathbf{p}_2 \rho_1 (m) \rho_2' (m') dm dm'
\over 4 \omega_{m} (\mathbf{p}_1)\omega_{m'} (\mathbf{2}_2)}
e^{-2\omega_{m} (\mathbf{p}_1)\tau_1   
-2 \omega_{m'} (\mathbf{p}_2)\tau_2}
\]
\beq
(
\omega_{m}(\mathbf{p}_1) - \omega_{m'}(\mathbf{p}_2)
- 
\omega_{m_1}(\mathbf{p}_1) - \omega_{m_2}(\mathbf{p}_2))
h_1 (m^2) h_2 (m^{\prime 2})
\psi_1 (\mathbf{p}_1) \psi_2 (\mathbf{p}_2) .
\label{f.10}
\eeq

This vanishes identically {\it because the only contribution to the $m$ and
$m'$ integrals comes from $m=m_1$ and $m'=m_2$} which leads to $
(\omega_{m}(\mathbf{p}_1) - \omega_{m'}(\mathbf{p}_2) -
\omega_{m_1}(\mathbf{p}_1) - \omega_{m_2}(\mathbf{p}_2))=0$.  The
disconnected term would not vanish without the $h(m^2)$ functions.  Instead
there would be non-zero terms that were independent of time (the real
time dependence cancels after performing the integrals over the
spatial coordinates).  This would lead to a violation of the Cook
condition.  In that case the strong limit would not exist.

The remaining terms are linear in the connected Euclidean four-point
Schwinger function.  The precise large time behavior depends on the form
of the connected four point function.  Following arguments in
\cite{jost}, we showed, \cite{Polyzou:2013nga} using a model
connected four-point function, the same $t^{-3/2}$ behavior
for the integrand in (\ref{e.10}) that one gets non-relativistically.
This is sufficient to satisfy the Cook condition.

\section{The $h$ function}

The key requirement in constructing a suitable injection operator is to 
construct an operator $h(\nabla^2)$ with the property 
\beq
h(\nabla_\x^2) f(\x) 
\label{g.1}
\eeq
has support for positive Euclidean time when $f(\x)$ has 
support for positive Euclidean time.  This support condition is 
clearly preserved when $h(\nabla^2)$ is a polynomial in
$\nabla^2$, however it is not obvious that $h(\nabla^2)$ preserves
the positive time support condition when $h(m^2)$ has
support on the real line.  The simplest counter example is the
unitary translation operator, which (1) can be formally represented as
an infinite series in derivatives and (2) changes (translates) the support
of functions.

In the unphysical case of a Lehmann weight with support on $N$ 
discrete masses an $h(x)$ that selects the $j^{th}$ mass 
has the form 
\beq
h_j (x) = \prod_{i\not=j}^N {x-m_i^2 \over m_j^2-m_i^2} .
\label{g.2}
\eeq
A more interesting, but still unphysical case, is when the support of
the Lehmann weight consists of a single mass (or several mass) and a
continuous spectrum with compact support, then the Weierstraass
approximation theorem implies $h(x)$ with the required
properties can uniformly approximated by a
polynomial on the support of the Lehmann weight.

The physically interesting case is when the support of the Lehmann
weight has some discrete masses and the support of the continuous part
extends from the lowest multiparticle threshold to infinity.  The
quantity of interest is
\[
( \psi , \Pi \Theta S_2 \Pi h(\nabla^2) \psi )
%\langle \psi \vert \Theta S_2 h(\nabla^2 ) \psi \rangle 
=
\]
\[
\int \psi (\mathbf{p},x_0) {e^{- \omega_m (\mathbf{p})(x_0 +y_0)}
\over 2 \omega_m (\mathbf{p}^2)} \rho(m^2) 
h({d^2 \over dy_0^2}- p^2)\psi (\mathbf{p},y_0)
d\mathbf{p} dx_0 dy_0 dm 
=  
\]
\beq
\int \psi (\mathbf{p},x_0) {e^{- \omega_m (\mathbf{p})(x_0 +y_0)}
\over 2 \omega_m (\mathbf{p}^2)} \rho(m^2) 
h(m^2)\psi (\mathbf{p},y_0)
d\mathbf{p} dx_0 dy_0 dm 
\label{g.3}
\eeq
where $\psi (\mathbf{p},x_0)$ is the partial Fourier transform 
with respect to the spatial variables of a positive-time support 
test function.
 
Our goal is to find states represented by positive time support
functions that asymptotically look like single-particle states.  The
support condition would be trivially satisfied if $h(\nabla^2)$ could
be approximated by a finite degree polynomial.  The problem is that
polynomials on semi-infinite intervals get large, however in this 
expression the polynomial growth is exponentially suppressed by
(see \ref{d.1a}):
\beq
{e^{- \omega_m (\mathbf{p})(x_0 +y_0)}
\over 2 \omega_m (\mathbf{p}^2)} \rho(m^2).
\label{g.4}
\eeq
If the polynomials in $m^2$ with respect to the weight (\ref{g.4}) are
complete, then it is possible to approximate the required
$h(\nabla^2)$ by a polynomial.  

The problem is to show that polynomials in $m^2$ with respect to the 
weight (\ref{g.4}) are complete.  Here it is enough to treat $\mathbf{p}$ as
a constant which is sufficient for three-momentum states with
compact support.
Note that the polynomials must be in 
$m^2$ rather than $m$ because the {\it square} of the mass is the Laplacian 
in this representation. 

Fortunately there is a sufficient condition for the a set of
polynomials with respect to a weight on a semi-infinite interval to be
complete.  The relevant condition, due to Carleman\cite{Carleman:1926}, 
depends on the growth of the Stiltjes moments
\beq
\gamma_n = \langle \psi  \vert \Theta S_2 
(\nabla^2)^n  \vert \psi \rangle =
\int_0^\infty \psi (\mathbf{p},x_0) {e^{- \omega_m (\mathbf{p})(x_0 +y_0)}
\over 2 \omega_m (\mathbf{p}^2)} \rho(m^2) 
m^{2n} \psi (\mathbf{p},y_0)
d\mathbf{p} dx_0 dy_0 dm . 
\label{g.5}
\eeq 
The Carleman condition states that if the following sum diverges 
\beq
\sum_{n=1}^\infty \gamma_n^{-{1 \over 2 n}} > \infty
\label{g.6}
\eeq
then the orthogonal polynomials in $m^2$ are complete
on the half line.   For this problem this means that 
functions $h(m^2)$ with the required properties can be 
approximated by a polynomial.

In the appendix we show that this condition is satisfied for
a polynomially bounded Lehmann weight $\rho (m)$.   

There remains the practical question of how to efficiently compute a suitable 
polynomial approximation to $h(m^2)$.  
%The simplest thing to try
%is to truncate the expansion of the Fourier representation of 
%$h$:
%\beq
%g(s) := {1 \over \sqrt{2 \pi}} 
%\int e^{-i m^2 s} h(m^2) 2m dm 
%\label{g.7}
%\eeq 
%and then construct a polynomial approximation by truncating the sum:
%\beq
%h_n(M^2) = {1 \over \sqrt{2\pi}} \sum_{k=0}^n \int g(s)  
%{(-is)^k \over k!}\nabla^{2k} .
%\label{g.8}
%\eeq
%The result is that by applying polynomial approximations 
%$f_n (x) := h_n(\nabla^2)f(x) $ gives a sequence of positive
%time support functions that converge to a mass eigenstate with 
%with the desired eigenvalue. 

\section{Computational methods in Euclidean space}

The discussion above implies that it is possible to formulate
the scattering problem directly in the Euclidean representation 
without analytic continuation.  While this does not mean that the
Euclidean representation is ideal from a computational point of view,
it is possible to reformulate the the asymptotic condition to 
take advantage of the Euclidean representation. 

The first simplification comes from using the invariance principle.
The invariance principle
\cite{kato:1966}\cite{Chandler:1976}\cite{simon}\cite{baumgartl:1983}
implies that if $f(E)$ is a function of bounded variation with a
positive derivative, $f'(E) >0$, then we can replace $H$ and $H_\alpha$
by $f(H)$ and $f(H_\alpha)$ in the asymptotic condition.  Using $f(x)
= - e^{-\beta x}$, where $\beta >0$ is a fixed positive constant, the
asymptotic condition 
\beq
\lim_{t \to \pm \infty}
\Vert e^{iHt } \vert f^{\pm} \rangle - 
J e^{-i H_0 t } \vert  f^0 \rangle \Vert =0,   
\label{h.0}
\eeq
is replaced by the equivalent asymptotic
condition
\beq
\lim_{s \to \pm \infty}
\Vert e^{ise^{-\beta H}} \vert f^{\pm} \rangle - 
J e^{i se^{-\beta H_0} } \vert  f^0 \rangle \Vert =0 .   
\label{h.1}
\eeq
In this expression the real time $t$ is replaced by the dimensionless 
quantity $s$.  With this substitution $f^0$ and $f^+$ remain 
unchanged.

Using unitarity of $e^{ie^{-\beta H}s}$ the asymptotic condition can be 
rewritten as 
\beq
\lim_{s \to \pm \infty}
\Vert  \vert f^{\pm} \rangle - 
e^{-ise^{-\beta H}} J e^{i se^{-\beta H_0} s} \vert f^0 \rangle \Vert =0   
\label{h.2}
\eeq
To compute 
\beq
\langle \x_1, \x_2 \vert 
e^{-ise^{-\beta H}} J e^{is e^{-\beta H_0}} \vert f^0 \rangle
\label{h.3}
\eeq
first note that $e^{-\beta H}$ has a spectrum between $[0,1]$ for 
$H \geq 0$.  Since this spectrum is compact it follows that
$e^{ise^{-\beta H}}$ can be uniformly approximated 
by a polynomial in
$e^{-\beta H}$:
\beq
e^{ise^{-\beta H}} = \sum_{n=0}^{\infty}  d_n e^{-\beta nH}
\label{h.4}
\eeq
The uniform convergence means that
\beq
\Vert\vert e^{ise^{-\beta H}} - \sum_{n=0}^N d_n e^{-\beta nH} \Vert\vert 
\leq \mbox{sup}_{x \in [-1,1]} \vert 
e^{isx} - \sum_{n=0}^N d_n x^{ n} \vert
\label{h.5}
\eeq
which can be evaluated by plotting the difference between the
functions.  While we have shown that this polynomial approximation can
be implemented efficiently (\cite{Kopp:2011vv}) for a wide range of
scattering energies, we show that if we use the injection
operator in section VI the polynomial can be summed to all orders.

Using the polynomial approximation in the expression for the 
initial and final scattering wave functions gives
\[
\langle \x_1, \x_2 \vert 
e^{-ise^{-\beta H}} J e^{is e^{-\beta H_0}}\vert f^0 \rangle \approx
\]
\[
\sum_{m=0}^N d_m \langle \mathbf{x}_1,\tau_{1}- m \beta  , 
\mathbf{x}_2,\tau_2- m\beta  \vert 
J e^{i se^{-\beta H_0} }\vert  f^0 \rangle  =   
\]
\[
\int \sum_{m=0}^N d_m
h_1 ({\partial^2 \over \partial \tau_1^2}- \mathbf{p}_1^2)
h_2 ({\partial^2 \over \partial \tau_2^2}- \mathbf{p}_2^2) 
\delta (\tau_{1}-\tau_{10}-m\beta) \delta (\tau_2 -\chi\tau_{10}-m \beta) 
{d \mathbf{p}_1 \over (2 \pi)^{3/2}}
{d \mathbf{p}_2 \over (2 \pi)^{3/2}}
\times 
\]
\[
e^{i \mathbf{p}_1 \cdot \mathbf{x}_1 + i \mathbf{p}_2 \cdot \mathbf{x}_2}
e^{is e^{-\beta (\omega_1 (\mathbf{p}_1)+\omega_2 (\mathbf{p}_2))}} 
\psi_1 (\mathbf{p}_1) \psi_2 (\mathbf{p}_2)  =
\]
\[
\int \sum_{m=0}^N d_m
h_1 (- {p}_1^2)
h_2 (- {p}_2^2)
{dp_{10} \over 2\pi}e^{i p_{10}(\tau_{1}-\tau_{10}-m\beta)}
{dp_{20} \over 2\pi}e^{i p_{20}(\tau_{2}-\chi \tau_{10}-m\beta)} 
{d \mathbf{p}_1 \over (2 \pi)^{3/2}}
{d \mathbf{p}_2 \over (2 \pi)^{3/2}}
\times 
\]
\beq
e^{i \mathbf{p}_1 \cdot \mathbf{x}_1 + i \mathbf{p}_2 \cdot \mathbf{x}_2}
e^{is e^{-\beta (\omega_1 (\mathbf{p}_1) + \omega_2 (\mathbf{p}_2))}} 
\psi_1 (\mathbf{p}_1) \psi_2 (\mathbf{p}_2)  
\label{h.6}
\eeq
We rewrite this in a form that allows a direct expression in terms of the
Fourier transform of the Schwinger function.
\[
\langle \x_1, \x_2 \vert 
e^{-ise^{-\beta H}} J e^{is e^{-\beta H_0}}\vert f^0 \rangle \approx
\]
\[
{1 \over (2\pi)^5} \int \sum_{m=0}^\infty d_m 
d^4p_1 d^4p_2 e^{i{p}_1 \cdot {x}_1 + i{p}_2 \cdot {x}_2}
e^{is e^{-\beta (\omega_1 (\mathbf{p}_1) + \omega_2 (\mathbf{p}_2))}} 
h_1 (- {p}_1^2)h_2 (- {p}_2^2)
\times
\]
\[
e^{-i p_{10}( \tau_{10}+m\beta) -i p_{20}(\chi \tau_{10}+m\beta)}
\psi_1 (\mathbf{p}_1) \psi_2 (\mathbf{p}_2)  =
\]
\[
{1 \over (2\pi)^5} \int  
d^4p_1 d^4p_2 e^{i{p}_1 \cdot {x}_1 + i{p}_22 \cdot {x}_2}
e^{is e^{-\beta (\omega_1 (\mathbf{p}_1) + \omega_2 (\mathbf{p}_2))}} 
h_1 (- {p}_1^2)h_2 (- {p}_2^2)
\times
\]
\beq
\sum_{m} d_m (e^{-i (p_{10}+ p_{20})\beta})^m
e^{-i p_{10}\tau_{10} -i p_{20}\chi \tau_{10})}
\psi_1 (\mathbf{p}_1) \psi_2 (\mathbf{p}_2)  
\label{h.17}
\eeq
In this form the polynomial can be summed explicitly to get 
\[
\langle \x_1, \x_2 \vert 
e^{-ise^{-\beta H}} J e^{is e^{-\beta H_0}}\vert f^0 \rangle =
\]
\[
{1 \over (2\pi)^5} \int
e^{i{p}_1 \cdot {x}_1 + i{p}_2 \cdot {x}_2}
e^{is e^{-\beta (\omega_1 (\mathbf{p}_1) + \omega_2 (\mathbf{p}_2))}} 
h_1 (- {p}_1^2)h_2 (- {p}_2^2)
\times
\]
\beq
e^{-ise^{-i\beta(p_{10}+ p_{20})}}
e^{-i p_{10}\tau_{10} -i p_{20}\chi \tau_{10})}
\psi_1 (\mathbf{p}_1) \psi_2 (\mathbf{p}_2)  
\label{h.8}
\eeq
Thus, by using the delta function in $\tau$ to define $J$ we can
replace the polynomial expansion by an integral.  Furthermore,
this is now exact.

Using these expressions in the formula for $S$-matrix elements gives
\[
\langle f^0_{f} \vert S \vert f^0_{i} \rangle = 
\langle f^+_{f} \vert f^-_{i} \rangle =
\]
\[
\int d^4 x_1' d^4 x_2' d^4 x_1 d^4 x_2 
{1 \over (2 \pi)^{10}}  
d^4p_1' d^4p_2' d^4p_1 d^4p_2 
\times
\]
\[
S_{4} (\theta x_2', \theta x_1', x_1, x_2)
h(-p_1^{\prime 2}) h(-p_2^{\prime 2}) 
e^{-i (p_1' \cdot x_1' + p_2' \cdot x_2' -  p_1 \cdot x_1 - p_2 \cdot x_2)}
e^{-is e^{-\beta (\omega_1' (\mathbf{p}'_1) + \omega'_2 (\mathbf{p}'_2))}} 
e^{2is e^{-i \beta (p_{10} + p_{20})}} 
\times
\]
\[
h(-p_1^2) h(-p_2^2) 
e^{-is e^{-\beta (\omega_1 (\mathbf{p}_1) + \omega_2 (\mathbf{p}_2))}} 
e^{i p_{10}'(\tau_{10}) i p_{20}'(\chi \tau_{10})}
e^{-i p_{10}(\chi \tau_{20}) -i p_{20}(\tau_{10})}
\times 
\]
\[
\psi_1^{\prime *} (\mathbf{p}_1') \psi_2^{\prime *} (\mathbf{p}_2')
\psi'_1 (\mathbf{p}_1) \psi_1' (\mathbf{p}_2) = 
\]
\[
\int 
{d^4p_1' d^4p_2' d^4p_2 d^4p_2 \over (2 \pi)^2}
\tilde{S}_{4} (-p_2', -p_1', p_1, p_2)
e^{-is e^{-i\beta (p_{10} + p_{20})}}
e^{is e^{-\beta (\omega_1 (\mathbf{p}_1') + \omega_2 (\mathbf{p}_2'))}} 
\times
\]
\[
e^{is e^{-\beta (\omega_1 (\mathbf{p}_1) + \omega_2 (\mathbf{p}_1))}} 
e^{-i p_{10}'(\tau_{10}) -i p_{20}'(\chi \tau_{10})}
e^{-i p_{10}(\tau_{10}) -i p_{20}(\chi\tau_{10})}
\times
\]
\beq
h(-p_1^{\prime 2}) h(-p_2^{\prime 2})
\psi_1' (\mathbf{p}_1') \psi'_2 (\mathbf{p}_2')
h(-p_1^2) h(-p_2^2)
\psi_1 (\mathbf{p}_1) \psi_2 (\mathbf{p}_2) = 
\label{h.9}
\eeq
One set of integrals can be performed due to the translational
invariance of the Green function.
The result is:
\[
\langle f^0_{f} \vert S \vert f^0_{i} \rangle = 
\]
\[
\langle \psi'_{1} \psi_2'  \vert S \vert \psi_1 \psi_2 \rangle = 
\]
\[
\lim_{s \to \infty}
\int
\psi_1^{\prime *} (\mathbf{p}_1') \psi_2^{\prime *}
(\mathbf{p}_2')
h_1'(-p_{1}^{\prime 2})
h_2'(-p_{2}^{\prime 2}) d4p_1'd4p_2'd^4p_1d^4p_2 \times
\]
\[
e^{-is e^{-\beta
(\omega'_1 (\mathbf{p}'_1) + \omega_2' (\mathbf{p}'_2))}}
e^{2is e^{-i\beta (p_{10} + p_{20})}} 
e^{-is e^{-\beta (\omega_1 (\mathbf{p}_1) + \omega_2 (\mathbf{p}_4))}}
\times 
\]
\beq 
e^{-i ((p_{10}+p_{10}')\tau_1 + (p_{20}+p_{20}')\chi \tau_2)} 
\tilde{S}_{4}(-p_2',-p_1',p_1,p_2)
h_1(-p_{1}^2)
h_2(-p_{2}^2) \psi_1 (\mathbf{p}_1) \psi_2
(\mathbf{p}_2)
\label{h.10}
\eeq
This leads to an expression for $S$ matrix elements as a 12 dimensional
integral.  While the form of this expression is complicated, it
involves the computation of a quadratic form involving the Fourier
transform of the Euclidean-four point functions without
analytic continuation.  The initial and final wave packets should be
narrow in 3-momentum, which will facilitate the computation of the
integrals.
It is still necessary to verify that
that the connected four point-function satisfies the Cook condition
and is consistent with reflection positivity. 

The sharp momentum transition matrix elements can be extracted
using sharply peaked $\psi_i(\mathbf{p})$:  
\beq
\langle \mathbf{p}_1', \mathbf{p}_2' \vert T \vert
\mathbf{p}_1, \mathbf{p}_2 \rangle =
{i \over 2 \pi}{\langle \psi_{1}'\psi_2' \vert \psi_1\psi_2 \rangle -
\langle \psi_1' \psi_2' \vert S \vert \psi_1 \psi_2 \rangle 
\over \langle \psi'_1\psi'_2 \vert \delta (E-H_0) \vert \psi_1 \psi_2'
\rangle >  
%( f_{0f}, J^{\dagger} \Theta G J \delta (E-H_0)  
%f_{0i})
}
\label{h.11}
\eeq

The calculation should converge as $s$ gets large.  The choice of
$\beta$ is in principle arbitrary, but it should be close to the 
inverse of the working energy scale.  
The parameter $\chi$ should be of order unity but larger than 1.  

\section{Summary}

The purpose of this paper was to demonstrate how to formulate the
scattering asymptotic condition using strong limits in the formulation
of quantum field theory defined by a collection of reflection-positive
Schwinger functions.  In this formalism the Hilbert space inner
product is expressed as a quadratic form with a kernel consisting of
the product of a Schwinger functions and a Euclidean time reversal on
the final states.  Hilbert space vectors have positive Euclidean
relative time support.

The main problem addressed in this paper was how to construct the
Euclidean analog of Haag-Ruelle quasilocal fields.  These are needed
to get a scattering theory that can be formulated in terms of strong
limits and to treat scattering where the initial and final particles
may be composite.  In order to achieve the desired result it was
necessary to show that certain polynomials are complete with respect
to a certain weight function.  This was established in section IX.
This result was needed to ensure that the application of complicated
differential operators that project positive relative time support
functions on the desired asymptotic states do not change these support
conditions.

To achieve this result we used the fact that there are normalizable
vectors in this representation of the Hilbert spare that are
proportional to delta functions in the Euclidean time variable.  A
bi-product of this observation is an expression for scattering matrix
elements (\ref{h.10}) involving only quadratures involving
sharp-momentum wave packets and the Fourier transform of the Schwinger
functions.  Previous work suggest that this method should be
applicable from low to intermediate energies.

\begin{acknowledgments}
The authors would like to thank Palle Jorgensen whose remarks
contributed materially to this work.  This work was performed under
the auspices of the U. S. Department of Energy, Office of Nuclear
Physics, under contract No. DE-FG02-86ER40286 with the University of
Iowa.
\end{acknowledgments}

\section{appendix} 

A sufficient condition for the orthogonal polynomials with respect
to a measure on $[0,\infty]$ to be complete is that the moments    
$\{\gamma_n\}$ of the measure satisfy the Carleman condition
\cite{Akheizer:1965}\cite{Carleman:1926}
\beq
\sum_{n=0}^\infty  \vert \gamma_n \vert^{-{1\over 2n}} > \infty
\label{ap:1}
\eeq
The moments of interest have the form
\beq
\gamma_n := \int_0^\infty {e^{- \sqrt{m^2 + \mathbf{p}^2} \tau} 
\over 2 \sqrt{m^2 + \mathbf{p}^2}} \rho(m) 
m^{2n} dm 
\label{ap:2}
\eeq
where 
$\tau = \tau_1 + \tau_2 >0$ and 
$\rho (m)$ is polynomially bounded.  Since $\rho (m)$ is polynomially
bounded we replace $\rho(m)$ in (\ref{ap:2}) by $m^k$.  
\beq
\gamma_n \to \gamma_n' = 
\int_0^\infty {e^{- \sqrt{m^2 + \mathbf{p}^2} \tau} 
\over 2 \sqrt{m^2 + \mathbf{p}^2}}  
m^{2n+k} dm .
\label{ap:3}
\eeq
If we make the substitutions $m = p \sinh (\eta)$ and 
$p \cosh (\eta) = \sqrt{m^2 + \mathbf{p}^2} $ this integral becomes
\beq
{1 \over 2} \int_0^\infty
{e^{-p\tau \cosh (\eta)} \over \cosh (\eta)} 
( p \sinh (\eta))^{2n+k} \cosh (\eta) d\eta
\label{ap:4}
\eeq
After the variable change $u=p \tau \cosh (\eta) $ this becomes 
\[
{1 \over 2} p^{2n+k} \int_{p\tau}^\infty e^{-u} 
({u^2 \over p^2 \tau^2}-1)^{{2n+k-1 \over 2}} {du \over p\tau} \leq
{1 \over 2} p^{2n+k} \int_{p\tau}^\infty e^{-u} 
({u \over p \tau})^{2n+k-1 } {du \over p\tau} \leq 
\]
\beq
{1 \over 2} p^{2n+k} \int_{0}^\infty e^{-u} 
({u \over p \tau})^{2n+k-1 } {du \over p\tau} =
{1 \over2} \tau^{-2n-k} \Gamma (2n+k-2)
\label{ap:5}
\eeq
using the the representation of the Gamma functions \cite{Abramowitz}
(equation 6.1.38):
\beq
\Gamma (x+1) = \sqrt{2 \pi} x^{x+1/2} e^{-x+\theta/12x}
\label{ap:6}
\eeq
where $\theta$ is a number between 0 and 1 that depends on $x$  gives
\beq
\gamma_n \leq 
\sqrt{\pi \over 2}\tau^{-2n-k} (2n+k-2)^{2n+k-3/2}e^{-(2n+k-2) +{\theta \over 2n+k-2}}
\label{ap:7}
\eeq
which gives
\beq
{1 \over \gamma_n } \geq 
\sqrt{2 \over pi}\tau^{2n+k} (2n+k-2)^{-2n-k+3/2}e^{(2n+k-2)-{\theta \over 2n+k-2}}
\label{ap:8}
\eeq
and 
\beq
({1 \over \gamma_n })^{{1\over 2n}} \geq 
\sqrt{2 \over \pi}^{1\over 2n}\tau^{1+k/2n} (2n+k-2)^{-1-(k-3/2)/2n}
e^{(1+(k-2)/2n)-{\theta \over 1+(k-2)/2n}}
\label{ap:9}
\eeq
For large $n$ the sum behaves like 
\beq
\sum {1 \over n +k/2-1} 
\label{ap:10}
\eeq
which diverges.  This imples that $h(\nabla^2)$ can be approximated
by a polynomial for any well behaved polynomially bounded Lehmann weight.

%\bibliography{master_bib_file.bib}

\begin{thebibliography}{25}
\expandafter\ifx\csname natexlab\endcsname\relax\def\natexlab#1{#1}\fi
\expandafter\ifx\csname bibnamefont\endcsname\relax
  \def\bibnamefont#1{#1}\fi
\expandafter\ifx\csname bibfnamefont\endcsname\relax
  \def\bibfnamefont#1{#1}\fi
\expandafter\ifx\csname citenamefont\endcsname\relax
  \def\citenamefont#1{#1}\fi
\expandafter\ifx\csname url\endcsname\relax
  \def\url#1{\texttt{#1}}\fi
\expandafter\ifx\csname urlprefix\endcsname\relax\def\urlprefix{URL }\fi
\providecommand{\bibinfo}[2]{#2}
\providecommand{\eprint}[2][]{\url{#2}}

\bibitem[{\citenamefont{Osterwalder and Schrader}(1973)}]{Osterwalder:1973dx}
\bibinfo{author}{\bibfnamefont{K.}~\bibnamefont{Osterwalder}} \bibnamefont{and}
  \bibinfo{author}{\bibfnamefont{R.}~\bibnamefont{Schrader}},
  \bibinfo{journal}{Commun. Math. Phys.} \textbf{\bibinfo{volume}{31}},
  \bibinfo{pages}{83} (\bibinfo{year}{1973}).

\bibitem[{\citenamefont{Schwinger}(1958)}]{Schwinger:pna}
\bibinfo{author}{\bibfnamefont{J.~S.} \bibnamefont{Schwinger}},
  \bibinfo{journal}{Proc. Natl. Acad. Sci. U. S.}
  \textbf{\bibinfo{volume}{44}}, \bibinfo{pages}{956} (\bibinfo{year}{1958}).

\bibitem[{\citenamefont{Glimm and Jaffe}(1981)}]{glimm:1981}
\bibinfo{author}{\bibfnamefont{J.}~\bibnamefont{Glimm}} \bibnamefont{and}
  \bibinfo{author}{\bibfnamefont{A.}~\bibnamefont{Jaffe}},
  \emph{\bibinfo{title}{Quantum Physics; A functional Integral Poinct of View}}
  (\bibinfo{publisher}{Springer-Verlag}, \bibinfo{year}{1981}).

\bibitem[{\citenamefont{Frohlich et~al.}(1983)\citenamefont{Frohlich,
  Osterwalder, and Seiler}}]{Frohlich:1983kp}
\bibinfo{author}{\bibfnamefont{J.}~\bibnamefont{Frohlich}},
  \bibinfo{author}{\bibfnamefont{K.}~\bibnamefont{Osterwalder}},
  \bibnamefont{and} \bibinfo{author}{\bibfnamefont{E.}~\bibnamefont{Seiler}},
  \bibinfo{journal}{Annals Math.} \textbf{\bibinfo{volume}{118}},
  \bibinfo{pages}{461} (\bibinfo{year}{1983}).

\bibitem[{\citenamefont{Reed and Simon}(1979)}]{simon}
\bibinfo{author}{\bibfnamefont{M.}~\bibnamefont{Reed}} \bibnamefont{and}
  \bibinfo{author}{\bibfnamefont{B.}~\bibnamefont{Simon}},
  \emph{\bibinfo{title}{Methods of Modern mathematical Physics}}, vol.
  \bibinfo{volume}{III Scattering Theory} (\bibinfo{publisher}{Academic Press},
  \bibinfo{year}{1979}).

\bibitem[{\citenamefont{Baumg\"artel and Wollenberg}(1983)}]{baumgartl:1983}
\bibinfo{author}{\bibfnamefont{H.}~\bibnamefont{Baumg\"artel}}
  \bibnamefont{and}
  \bibinfo{author}{\bibfnamefont{M.}~\bibnamefont{Wollenberg}},
  \emph{\bibinfo{title}{Mathematical Scattering Theory}}
  (\bibinfo{publisher}{Spinger-Verlag, Berlin}, \bibinfo{year}{1983}).

\bibitem[{\citenamefont{Wessels and Polyzou}(2004)}]{Wessels:2003af}
\bibinfo{author}{\bibfnamefont{V.}~\bibnamefont{Wessels}} \bibnamefont{and}
  \bibinfo{author}{\bibfnamefont{W.}~\bibnamefont{Polyzou}},
  \bibinfo{journal}{Few Body Syst.} \textbf{\bibinfo{volume}{35}},
  \bibinfo{pages}{51} (\bibinfo{year}{2004}), \eprint{nucl-th/0312004}.

\bibitem[{\citenamefont{Kopp and Polyzou}(2012)}]{Kopp:2011vv}
\bibinfo{author}{\bibfnamefont{P.}~\bibnamefont{Kopp}} \bibnamefont{and}
  \bibinfo{author}{\bibfnamefont{W.}~\bibnamefont{Polyzou}},
  \bibinfo{journal}{Phys.Rev.} \textbf{\bibinfo{volume}{D85}},
  \bibinfo{pages}{016004} (\bibinfo{year}{2012}), \eprint{1106.4086}.

\bibitem[{\citenamefont{Polyzou}(2014)}]{Polyzou:2013nga}
\bibinfo{author}{\bibfnamefont{W.}~\bibnamefont{Polyzou}},
  \bibinfo{journal}{Phys.Rev.} \textbf{\bibinfo{volume}{D89}},
  \bibinfo{pages}{076008} (\bibinfo{year}{2014}), \eprint{1312.3585}.

\bibitem[{\citenamefont{Cook}(1957)}]{Cook:1957}
\bibinfo{author}{\bibfnamefont{J.}~\bibnamefont{Cook}}, \bibinfo{journal}{J.
  Math. Phys.} \textbf{\bibinfo{volume}{36}}, \bibinfo{pages}{82}
  (\bibinfo{year}{1957}).

\bibitem[{\citenamefont{Haag}(1958)}]{Haag:1958vt}
\bibinfo{author}{\bibfnamefont{R.}~\bibnamefont{Haag}}, \bibinfo{journal}{Phys.
  Rev.} \textbf{\bibinfo{volume}{112}}, \bibinfo{pages}{669}
  (\bibinfo{year}{1958}).

\bibitem[{\citenamefont{Brenig and Haag}(1959)}]{Brenig:1959}
\bibinfo{author}{\bibfnamefont{W.}~\bibnamefont{Brenig}} \bibnamefont{and}
  \bibinfo{author}{\bibfnamefont{R.}~\bibnamefont{Haag}},
  \bibinfo{journal}{Fort. der Physik} \textbf{\bibinfo{volume}{7}},
  \bibinfo{pages}{183} (\bibinfo{year}{1959}).

\bibitem[{\citenamefont{Ruelle}(1962)}]{Ruelle:1962}
\bibinfo{author}{\bibfnamefont{D.}~\bibnamefont{Ruelle}},
  \bibinfo{journal}{Helv. Phys. Acta.} \textbf{\bibinfo{volume}{35}},
  \bibinfo{pages}{147} (\bibinfo{year}{1962}).

\bibitem[{\citenamefont{Jost}(1965)}]{jost}
\bibinfo{author}{\bibfnamefont{R.}~\bibnamefont{Jost}},
  \emph{\bibinfo{title}{The General Theory of Quantized Fields}}
  (\bibinfo{publisher}{AMS}, \bibinfo{year}{1965}).

\bibitem[{\citenamefont{Klein and L.}(1981)}]{Klein:1981}
\bibinfo{author}{\bibfnamefont{A.}~\bibnamefont{Klein}} \bibnamefont{and}
  \bibinfo{author}{\bibfnamefont{L.}~\bibnamefont{L.}}, \bibinfo{journal}{J.
  Functional Anal.} \textbf{\bibinfo{volume}{44}}, \bibinfo{pages}{121}
  (\bibinfo{year}{1981}).

\bibitem[{\citenamefont{Klein and L.}(1983)}]{Klein:1983}
\bibinfo{author}{\bibfnamefont{A.}~\bibnamefont{Klein}} \bibnamefont{and}
  \bibinfo{author}{\bibfnamefont{L.}~\bibnamefont{L.}}, \bibinfo{journal}{Comm.
  Math. Phys} \textbf{\bibinfo{volume}{87}}, \bibinfo{pages}{469}
  (\bibinfo{year}{1983}).

\bibitem[{\citenamefont{Coester}(1965)}]{Coester:1965zz}
\bibinfo{author}{\bibfnamefont{F.}~\bibnamefont{Coester}},
  \bibinfo{journal}{Helv. Phys. Acta} \textbf{\bibinfo{volume}{38}},
  \bibinfo{pages}{7} (\bibinfo{year}{1965}).

\bibitem[{\citenamefont{Chandler and Gibson}(1980)}]{Chandler:1980}
\bibinfo{author}{\bibfnamefont{C.}~\bibnamefont{Chandler}} \bibnamefont{and}
  \bibinfo{author}{\bibfnamefont{A.}~\bibnamefont{Gibson}},
  \bibinfo{journal}{Mathematical Methods and Applications of Scattering Theory,
  Lecture Notes in Physics.} \textbf{\bibinfo{volume}{130}},
  \bibinfo{pages}{134} (\bibinfo{year}{1980}).

\bibitem[{\citenamefont{Coester and Polyzou}(1982)}]{Coester:1982vt}
\bibinfo{author}{\bibfnamefont{F.}~\bibnamefont{Coester}} \bibnamefont{and}
  \bibinfo{author}{\bibfnamefont{W.~N.} \bibnamefont{Polyzou}},
  \bibinfo{journal}{Phys. Rev.} \textbf{\bibinfo{volume}{D26}},
  \bibinfo{pages}{1348} (\bibinfo{year}{1982}).

\bibitem[{\citenamefont{Keister and Polyzou}(1991)}]{Keister:1991sb}
\bibinfo{author}{\bibfnamefont{B.~D.} \bibnamefont{Keister}} \bibnamefont{and}
  \bibinfo{author}{\bibfnamefont{W.~N.} \bibnamefont{Polyzou}},
  \bibinfo{journal}{Adv. Nucl. Phys.} \textbf{\bibinfo{volume}{20}},
  \bibinfo{pages}{225} (\bibinfo{year}{1991}).

\bibitem[{\citenamefont{Carleman}(1926)}]{Carleman:1926}
\bibinfo{author}{\bibfnamefont{T.}~\bibnamefont{Carleman}},
  \emph{\bibinfo{title}{Les fonctions quasi analytiques, Collection de
  Monographies sur la Th\'eorie des Fonctions}}
  (\bibinfo{publisher}{Gauthier--Villars, Paris}, \bibinfo{year}{1926}).

\bibitem[{\citenamefont{Kato}(1966)}]{kato:1966}
\bibinfo{author}{\bibfnamefont{T.}~\bibnamefont{Kato}},
  \emph{\bibinfo{title}{Perturbation theory for linear operators}}
  (\bibinfo{publisher}{Spinger-Verlag, Berlin}, \bibinfo{year}{1966}).

\bibitem[{\citenamefont{Chandler and Gibson}(1976)}]{Chandler:1976}
\bibinfo{author}{\bibfnamefont{C.}~\bibnamefont{Chandler}} \bibnamefont{and}
  \bibinfo{author}{\bibfnamefont{A.}~\bibnamefont{Gibson}},
  \bibinfo{journal}{Indiana Journal of Mathematics.}
  \textbf{\bibinfo{volume}{25}}, \bibinfo{pages}{443} (\bibinfo{year}{1976}).

\bibitem[{\citenamefont{Akheizer}(1965)}]{Akheizer:1965}
\bibinfo{author}{\bibfnamefont{N.~I.} \bibnamefont{Akheizer}},
  \emph{\bibinfo{title}{The Classical Moment Problem and some related questions
  in analysis}} (\bibinfo{publisher}{Oliver and Boyd, Edinburgh and London},
  \bibinfo{year}{1965}).

\bibitem[{\citenamefont{Abramowitz and Stegun}(1972)}]{Abramowitz}
\bibinfo{author}{\bibfnamefont{M.}~\bibnamefont{Abramowitz}} \bibnamefont{and}
  \bibinfo{author}{\bibfnamefont{I.}~\bibnamefont{Stegun}},
  \emph{\bibinfo{title}{Handbook of Mathematical Functions with Formulas,
  Graphs, and Mathematical Tables}} (\bibinfo{publisher}{National Brueau of
  Standards}, \bibinfo{year}{1972}).

\end{thebibliography}
\end{document}